# AN EFFECTIVE AND SECURE USER AUTHENTICATED PROTOCOL FOR LOCATION SERVICES IN ROAD NETWORKS

Imran memon1 ,Hina memon[2]
[1]College of computer science ,Zhejiang university, China
[2]Institute of Mathematics and computer science, university of Sindh Jamshoro ,Pakistan

**Abstract**:
Road network is employed for exchanging the information among the vehicles where accidents and traffic information can be delivered, or receive services by an infrastructure. Although wireless communication systems yield an efficient traffic system and provide ease to the drivers, there is a chance of a traffic disturbance and risk to drivers through malicious information. So, there should be examined a way to limit the chance of an intermediate attack. This paper presents an efficient user based authentication protocol for location based services to secure address configuration for IPv6-based mix-zones over the road network. This protocol authenticates to inspect vehicles actions confidentially and have the following characteristics (1) Anonymous authentication: a message issuer can be authenticated. (2) Privacy: Communication content is confidential. The cost must be reduced through the address configuration scheme to improve the scalability. (3) Efficiency: it attains quick message verification, low storage requirements, and in case of a dispute, provides cost efficient identity tracking. Vehicles movement, the variation of velocity and distance are considered to maintain as many common users as possible by reducing the cost. The performance evaluation and cost analysis show that our framework can reduce the cost and gain outperformed results. This model can achieve reliability and efficiency with packet rate information. This user authenticated key establishment protocol has comparatively shorter time response, diminishes cost, less packet loss information and enhanced privacy preservation against malicious attacks compared with existing methods.

## 1. Introduction

The rapid development of vehicular technologies may apply new information and interconnection to improve the safety via vehicle to vehicle (V2V) and vehicle to infrastructure (V2I) communications [1]. Vehicular networks are also called Vehicular Ad Hoc Networks (VANETs), which are mostly used in Intelligent Transportation Systems (ITS) applications. In vehicular networks, the vehicles communicate with each other such as Inter-Vehicle Communication (IVC) and also with roadside base stations through a roadside unit. The vehicular networks provide safety for the users on the roads by providing timely information to the drivers. The vehicular network is a sub-class of mobile ad hoc networks (MANETs) and works on the same principle of mobile ad hoc networks (MANETs). In the early stage of 2000, MANET were one to one application but now a days VANETs have grown up in terms of inter-vehicle communication. VANET supports wide range of applications such as multi-hop message broadcasting over long distance and many other technologies that might uses UMTS, LTE, or WiMAX IEEE 802.16. For short range communication, it may use WLAN (either standard Wi-Fi), Bluetooth, Visible Light Communication and Infrared. Routing protocols in VANETs are significantly different from road networks, because various applications in VANET may have different QOS requirements for safe application. Broadcast routing in VANET is unlike from routing in road network due to various reasons such as rapidly changing network topology, wide range communication, and traffic pattern in different time and places. This may imply that conventional routing protocols for road network are not appropriate for most vehicular broadcast applications and they may also have different application requirements as compared to road network such as Infotainment applications, assistance co-operative awareness and Traffic efficiency management.

In VANETs each message need not to be verified and sent to the main server while on the other hand, in road networks every message must have to pass from main server and road side unit. These periodically broadcast messages are known as beacon messages. The content of beacon messages may include a vehicle's current position, velocity, and headway route. All the beacon messages have common requirements which are periodic broadcast and low latency. The beacon messages are useful for safety applications, for instance driver assistance, collision avoidance, and cruise control, etc. These applications require timely and correct information, and the typical beacon broadcast frequency, that might be in the range of 5-10 Hz [2]. The National Highway Trace Safety

Administration (NHTSA) has identified and set down requirements for Intelligent Vehicle Safety Applications [3]. Also, the work that would require V2V devices in latest vehicles in future has already been started on a regulatory proposal. In contrast, in V2I architecture the mobile vehicles directly connect with infrastructures set up along roads for sending and disseminating packets [4].

In Mix-zone server, two classical address configuration protocols such as stateful protocol [5] and other is stateless protocol established on duplicate address detection [6-10]. To consider latency and high cost, such protocols could not efficiently work in multi-hops network and road network environment. To reduce the delay and cost and to achieve address configuration in the road networks is an important challenge. Various messages for road conditions, congestion avoidance and detour notification for road authorization can spread by VANETs [11-17]. The value-added services and traffic associated message delivered by road network are used to improve drivers' wayfaring capability, toll payment services and provide internet access navigation etc. To address above challenges, this manuscript has proposed an effective protocol and anonymous authentication scheme for road network. Our scheme has significant feature that may compared with existing methods; (i) An anonymous authentication, it provides content secrecy communication (ii) It accomplishes low storage requirements.

Following are our main contributions in this work:

(1) We proposed location based routing protocol in road network to send data rapidly from the source vehicle to destination vehicle. A road network is an esteemed number of network gaps such as shortest paths, the average speed of the vehicles and the expected delay to transmit the data from one junction to another.

(2) We establish a privacy preserving authentication protocol to verify the vehicle activities in a privacy preserving manner. We proposed a new method to adjust the vehicles speed which reduces the vehicle delay that might suffer from the network gap problem.

(3) We proposed a novel vehicle orientated privacy preserving technique for road network services that is also proficient of protecting the Road side unit from attacks.

(4) Vehicles moving trend, distance difference and velocity difference, are considered to maintain as many common users as possible to reduce the cost. The performance evaluation and cost analysis indicate that our framework can reduce the cost factor and gain good performance. With roadside unit server assistance, the proposed model can perform reliably.

(5) The road side unit can immediately decide which messages are trustworthy, and then send its opinion to neighbouring vehicles immediately. With the aid of road side unit server, vehicles can successfully increase decision accuracy while making decisions about the event messages.

(6) The evaluation experiments based on NS-3 [18] to improve user's authenticated key establishment protocol , that have relatively shorter response time, reduce cost, less packet lost information and enhanced privacy preservation as compared with the proposed methods in [19] and [20].

Memon et al. [4] presented secure and effective Communication design with authenticated key establishment protocol for road networks. This proposed scheme is based on Mix-zone server ($MZ_s$) protocol and have achieved address configuration for users over the road network. Moreover, this scheme has achieved user communication security via authentication kind of road network. Considering some of these features; this work is totally different from our previous work [4] in terms of following aspects:

The system architectures are unlike our previous work. A road network is a special type of vehicular network, when a vehicle enters a new IP domain; its network prefix correspondingly changes. Therefore, the architecture in this scheme is based on multiple IP dynamics mix-zone server. In our previous work, a road network is limited to one IP domain, so the network prefix of a mobile node keeps unchanged. The previous work based upon One IP domain.

The IPV6 address structure has changed. In this scheme, the network architecture is based on road segments and various junction concepts. In order to improve the performance and contribute to the hierarchical routing which is made up of RS ID and node ID. In previous work, the network architecture is based on the MIX-ZONE SERVER tree topology where a gateway node is the root. Correspondingly, an address consists of three parts that might be the global routing prefix, the gateway node ID and the node ID. This protocol simultaneously achieves the address configuration and address reclamation at the same time, therefore the extra cost and delay caused by the address reclamation is avoided.

## 2. Address configuration protocol

### *2.1. Protocol's Structure*

The road network consists of mobile vehicles and RSU (road side unit) in this protocol [19-26]

communicating with Mix-zone server Internet over road side unit is shown in Fig. 1.

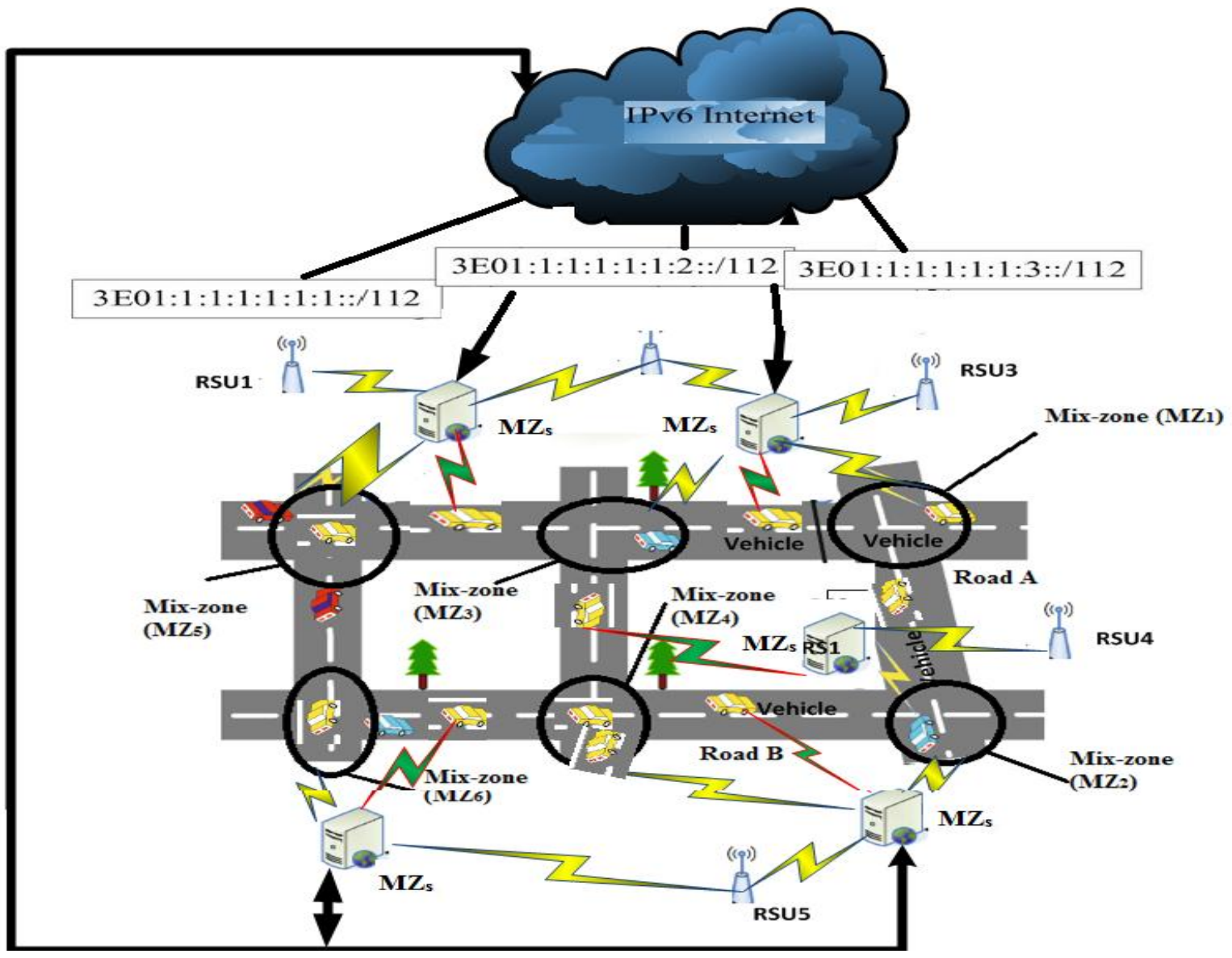

**Fig. 1.** Road network architecture.

### *2.2. Address Structure and address configuration algorithm*

The address structure for road network based on the proposed architecture is presented in Table 1 and Table 2.

**Table. 1** IPv6 address structure

| (128-*i*)bits | *i* bits |
|---|---|
| RSU ID | Mobile vehicle ID |

**Table. 2** Notations

| Notations | Descriptions | Notations | Descriptions |
|---|---|---|---|
| $U_i$ | Mobile vehicle | $p$ | a large prime |
| $MZ_s$ | Mix zone server | $\|\|$ | string concatenation |
| RSUj | Road side unit | $\oplus$ | XOR operation |
| $ID_i$ | identity of mobile vehicle $U_i$ | $\rightarrow$ | message transmission |
| RSUIDj | identity of Road side unit | $pw_i$ | password of $U_i$ |
| $r$ | master secret of $MZ_s$ | $h()$ | secure hash function |
| ST | Secret key between users and Mix-zones | s | server's secret key |

#### *2.2.1 Registration state*

In registration state, as the user $U_i$ wants to register with the $MZ_s$, it performs two steps with $MZ_s$.

Step:1 $\quad U_i \rightarrow MZ_s : (IDi, h(PWi), h(\cdot))$

The user $U_i$ after freely choosing its password $pw_i$ and identity $ID_i$ performs an iris scan with a capture and generates a template $MZ_s$ that describes the iris template of $U_i$. . This protocol adopts iris as a characteristic since it is harder to be compromised in contrast to their characteristic information such as fingerprint. Then, the user $U_i$ selects a secure one-way hash function $h(\cdot): \{0,1\}^* \rightarrow \{0,1\}^k$ and calculates $h(pwi)$, then it submits $\{ID, ST, h(pwi), h(\cdot)\}$ to the $MZ_s$ through a secure channel.

Step:2 $\quad MZ_s U_i \rightarrow U : (ID_i, h(PW_i), (\cdot))$

After getting the information from the user $U_i$ , $MZ_s$ calculates the secret information and encrypts $ID_i$ and ST via the server's secret key s to attain $T = E_s(ID \| ST)$ . Then, it conveys this smart card to the user $U_i$ in a safe method in fig.2. The user $U_i$ preserves the password $pw_i$ , the identity ID and the smart card confidentially for the authentication manner Case 1. Logging in for the first time, as shown in Fig.3 and Fig.4. Case 2. On a following login, as shown in Fig.5 and Fig.6.

### *2.2.2. First time login state*

Step:1 $Qj1 = T_{rj}(x) \bmod p,\ Qj2 = T_{rj}(Q) \bmod p,$

$Y1 = h(Qj2),\ Y2 = h(Qj1 \| Qj2 \| Qi1 \| Y1 \| X2 \| Bj).$

Step:2 $MZ_s \rightarrow RSUj\ \{Qi1\ X1, X2, Qj1, Y1, Y2\}$

$MZ_s \rightarrow$ *sends* $\{Qi1, X1, X2, Qj1, Y1, Y2\}$ *to the RSUj* $(RSUj)$.

### *2.2.3. First-time login Authentication and key agreement state*

Step:1 $\{Qi1, X1, X2, Qj1, Y1, Y2\},$

The $MZ_s$ computes

$Qi2^* \equiv T_r(Q_{i1}) \bmod p, IDi^* = h(Qi2^*) \oplus X1, Ai^* = h(IDi^* \| r),$

$Qj2^* \equiv T_r(Q_{j1}) \bmod p, RSUIDj^* = h(Qj2^*) \oplus Y1, Bj^* = h(RSUIDj^*),$

Step 2:
The $MZ_s$ checks if

$X2 = h(Ai^* \| RSUIDj \| Qi1 \| Qi2^* \| X1), Y2 = h(Qj1 \| Qj2^* \| Qi1 \| Y1 \| X2 \| Bj^*).$

$Zi = h(Qi1 \| Qj1 \| RSUIDj^* \| Ai^* \| X2), Zj = h(Bj^* \| Qi1 \| Qj1 \| Zi).$

Step:3 $MZ_s \rightarrow RSUj:\ \{Zi, Zj\}$

Step:4 $RSUj \rightarrow MZ_s:\ \{Qj1, Rj, CIDi, t, R\}$

Step:5
The smart card first computes:

$Zi^* = h(Qi1 \| Qj1 \| RSUIDj \| Ai \| X2)$

and checks whether

$Rj = h(Zi^* \| T_{r_i}(Q_{j1}) \| CIDi \| t \| R)$

$Ri = h(RSUIDj \| Qi1 \| Qj1 \| V \| T_{r_i}(Q_{j1})), d = h(RSUIDj \| pwi \| IDi) \oplus V$

and stores

$\{RSUIDj, d, CIDi, t\}$

$SK = h(T_{r_i}(Q_{j1}) \| RSUIDj)$

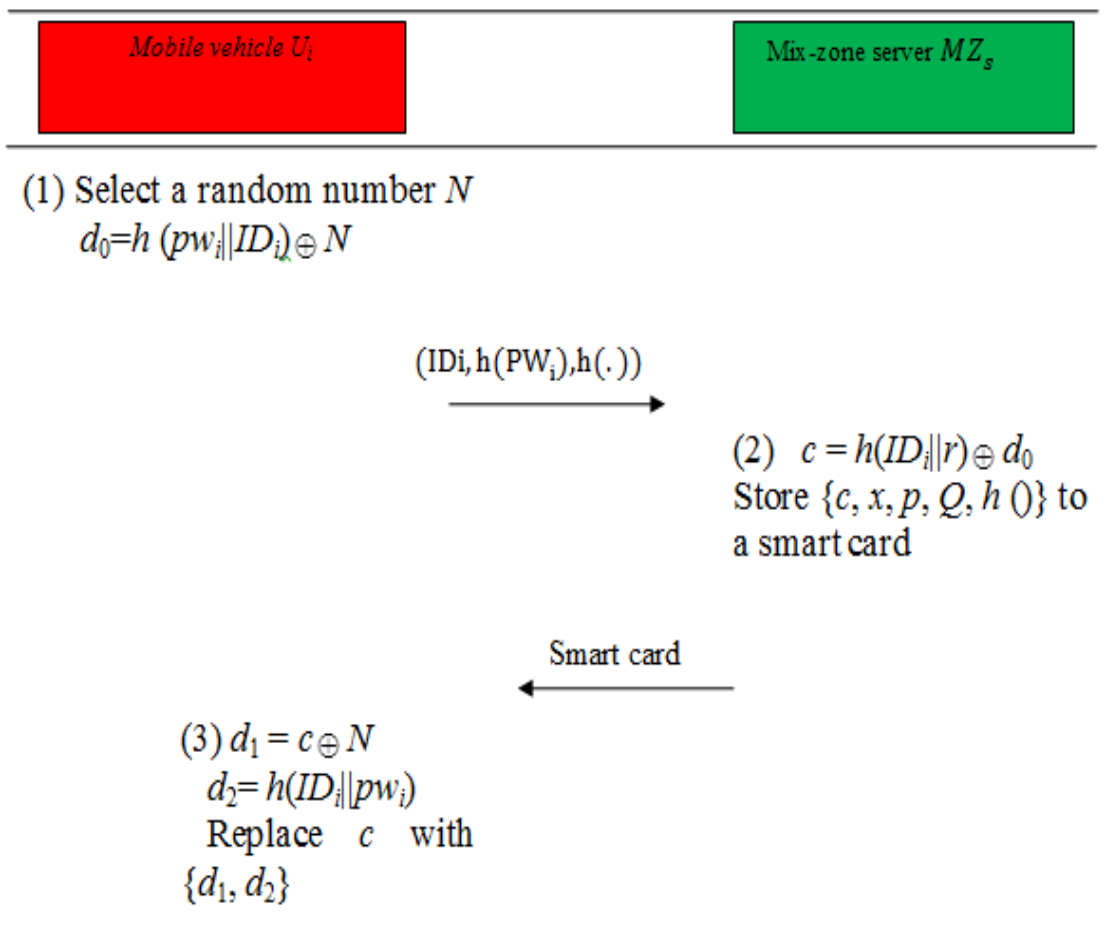

**Fig. 2.** Registration state.

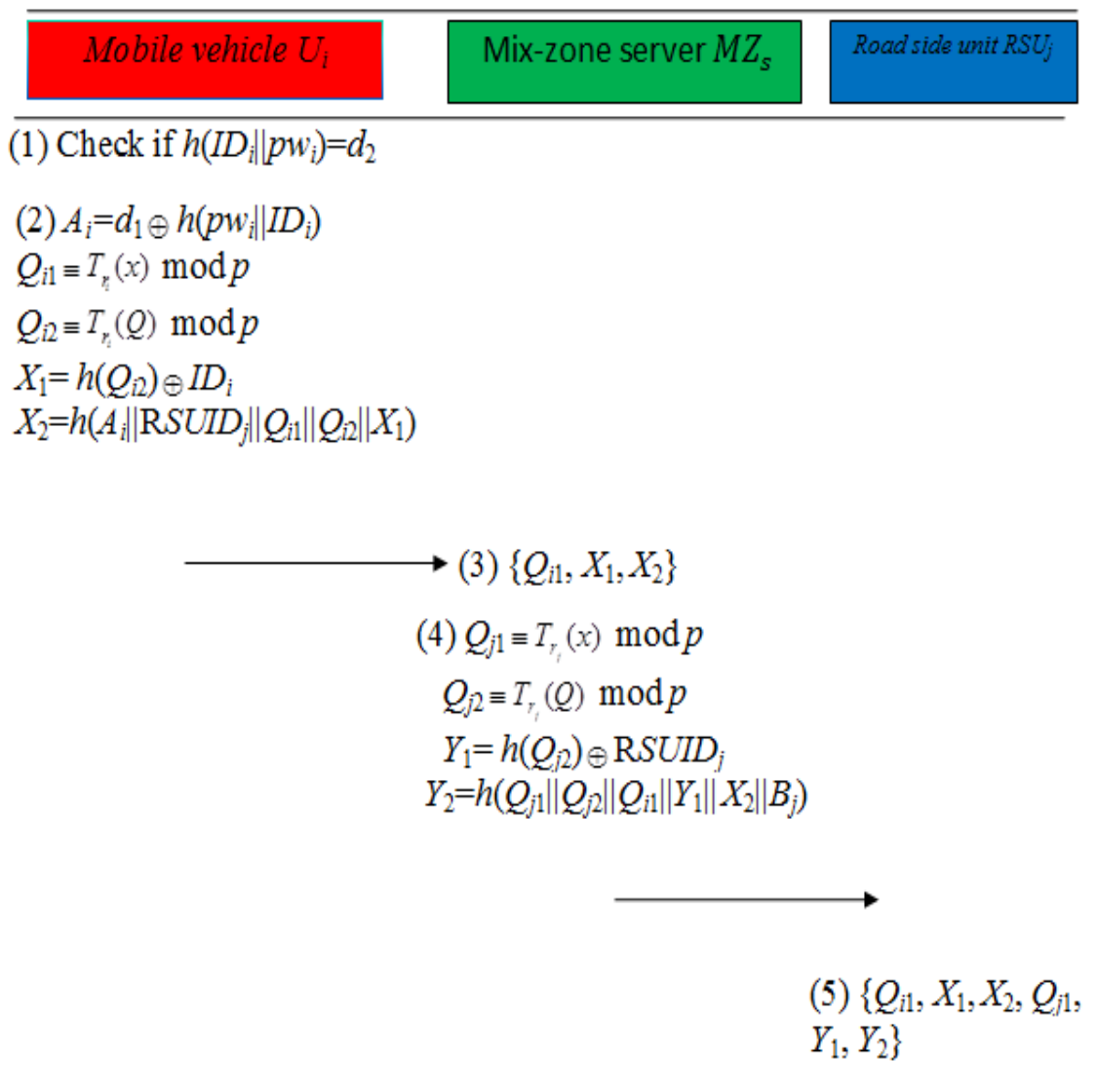

**Fig. 3.** First time login state.

Step:7 $U_i \rightarrow MZ_s : \{MZ_s\}$

The card sends

$\{MZ_s\}$ to the $RSUj$.

Step:8 $Ri = h(RUSIDj \| Qi1 \| Qj1 \| h(CIDi \| Bj \| t) \| T_{r_j}(Q_{i1}))$

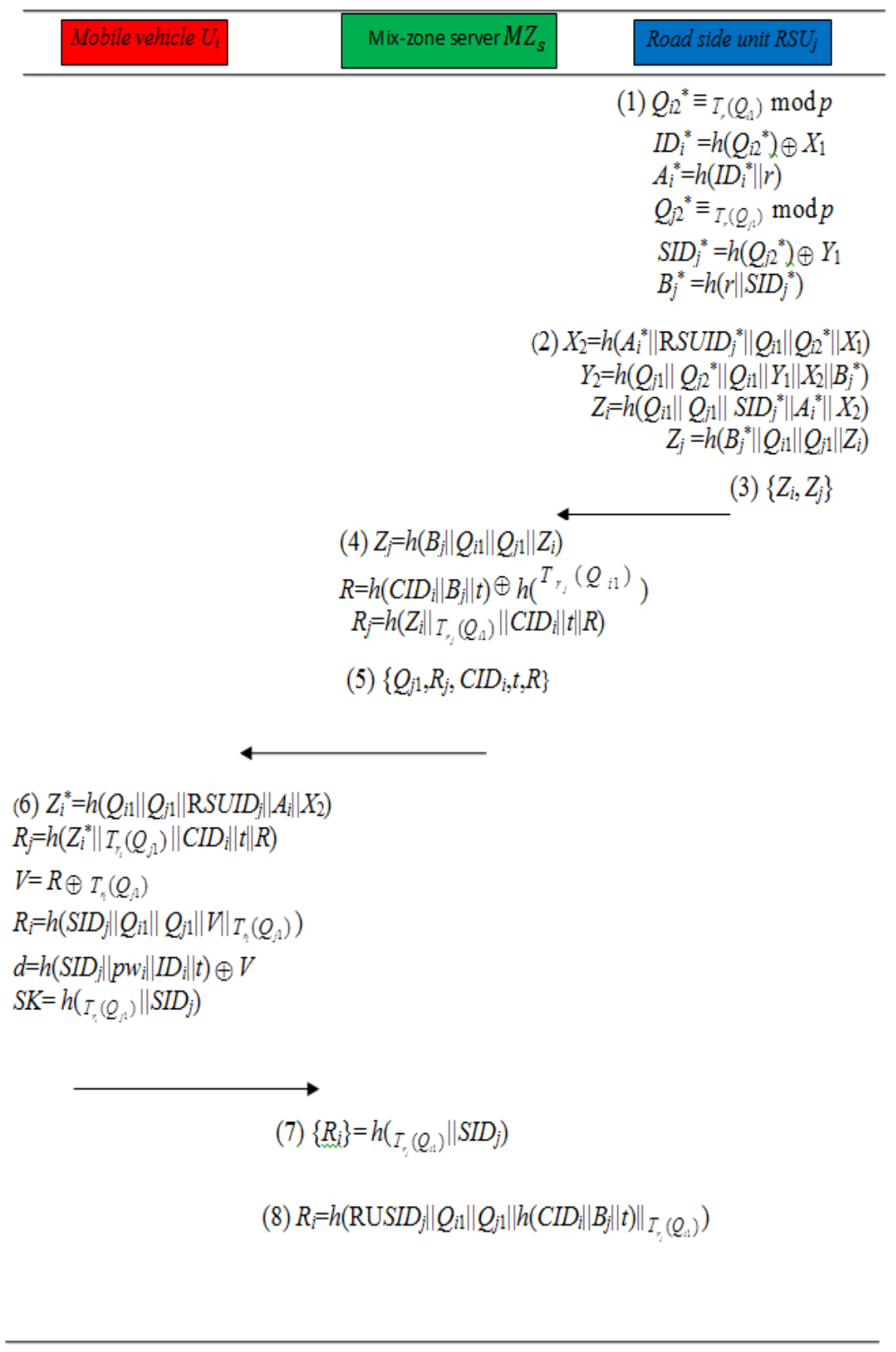

**Fig. 4**. First-time login Authentication and key agreement state

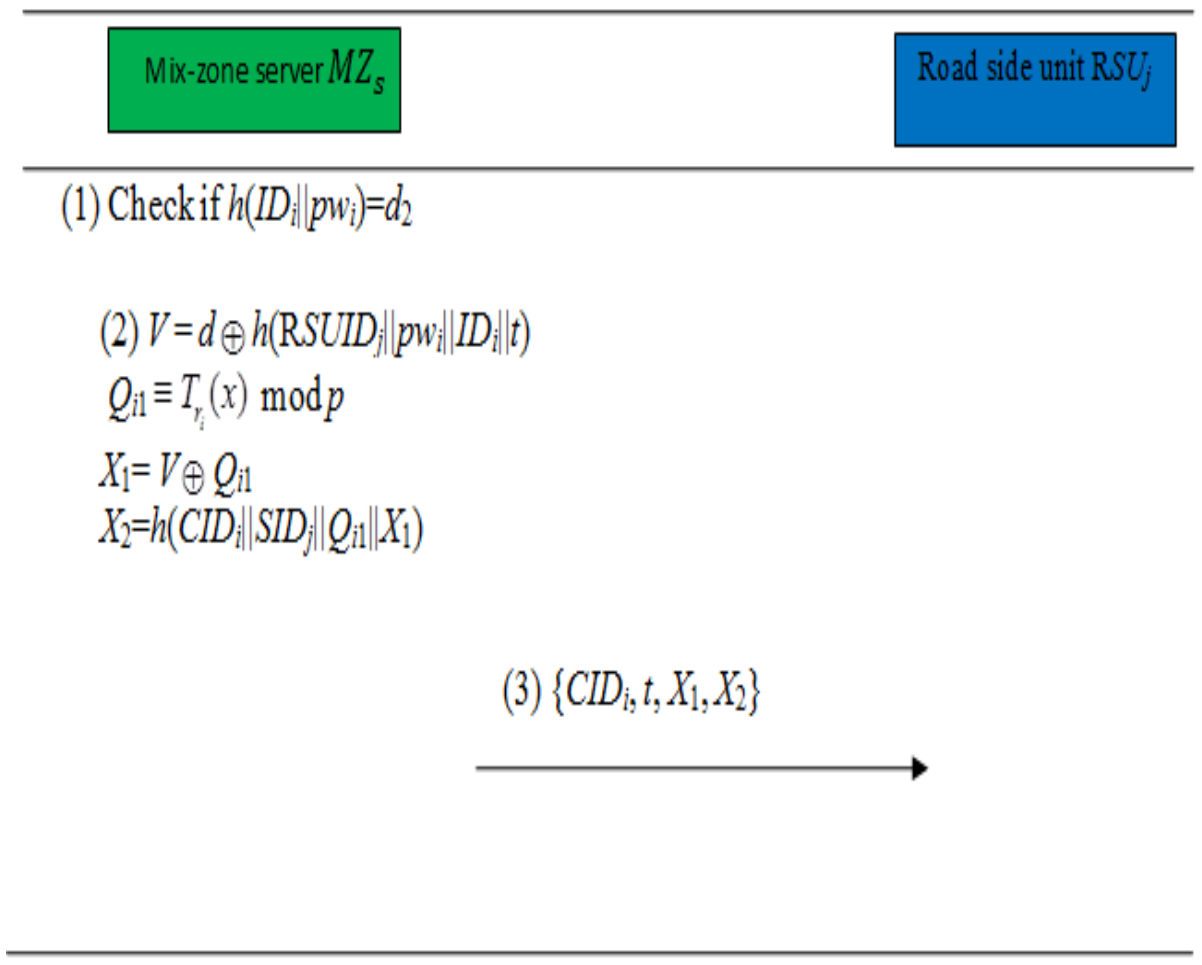

**Fig. 5.** Consequent login state

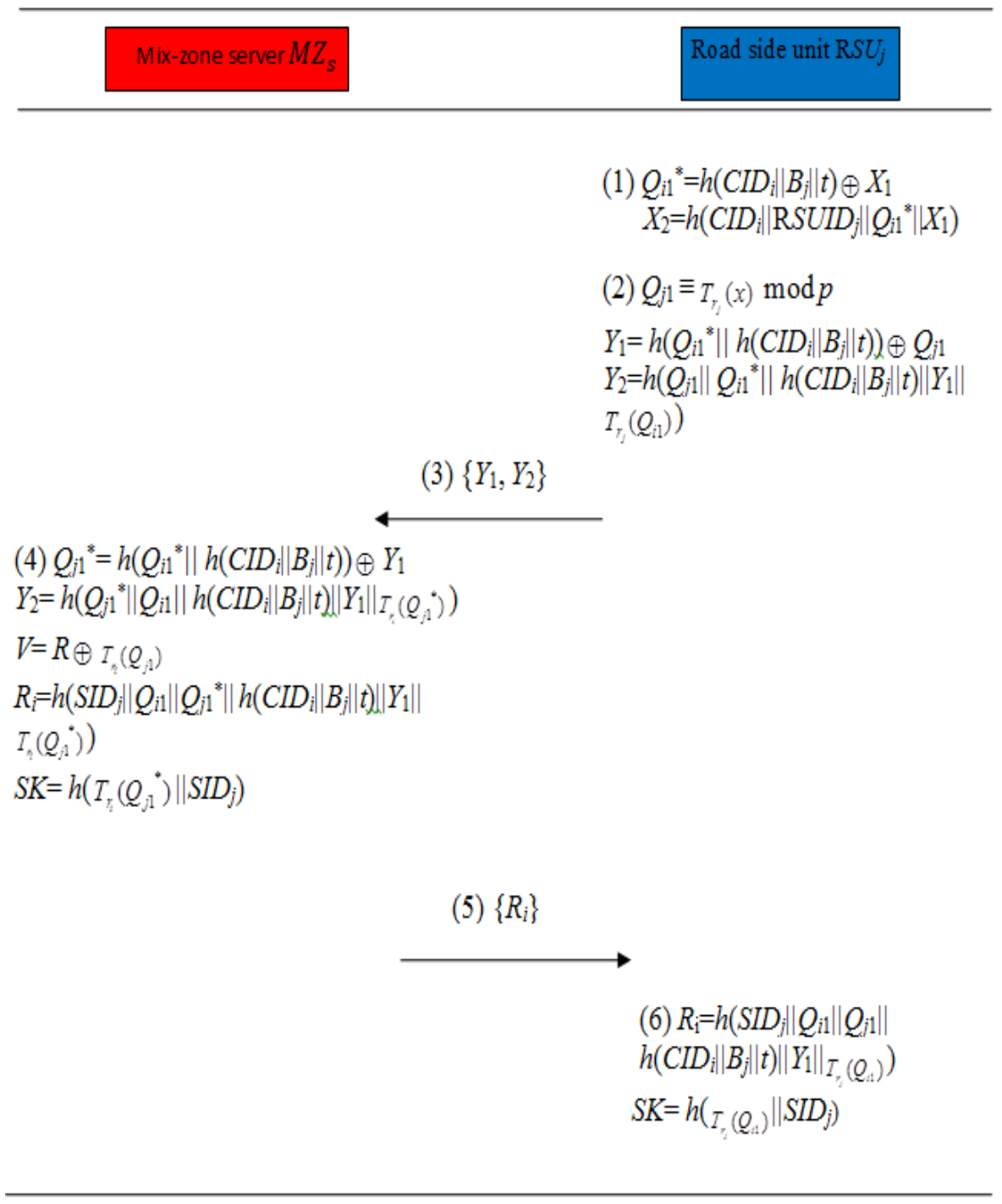

**Fig. 6**. Consequent login Authentication and key agreement state

### *2.2.4. Consequent login Authentication and key agreement state*

Step 1. On the reception of request message $\{CID_i, t, X_1, X_2\}, MZ_s$ primarily checks if $t$ is valid. Then, if it is valid, $MZ_s$ computes $Q_{i1}^* = h(CID_i || B_j || t) \oplus X_1$. Then $RSU_j$ checks whether

$$X_2 = h(CID_i || RSUID_j || Q_{i1}^* || X_1).$$

Step:2 $\quad Q_{j1} \equiv T_{r_j}(x) mod p, Y_1 = h(Q_{i1}^* || h(CID_i || B_j || t)) \oplus Q_{j1},$

$$Y_2 = h(Q_{j1} || Q_{i1}^* || h(CID_i || B_j || t) || Y_1 || T_{r_j}(Q_{i1})).$$

Step:3 $\quad RSU_j \rightarrow MZ_s : \{Y_1, Y_2\}$

Step:4 The smart card initially calculates

$$Q_{j1}^* = h(Q_{i1}^* || h(CID_i || B_j || t)) \oplus Y_1$$

and checks whether

$$Y_2 = h(Q_{j1}^* || Q_{i1} || h(CID_i || B_j || t) || Y_1 || T_{r_i}(Q_{j1}^*))$$

If case of above equality, the smart card calculates $V = R \oplus T_{r_i}(Q_{j1})$,

$$R_i = h(RSUID_j || Q_{i1} || Q_{j1}^* || h(CID_i || B_j || t) || Y_1 || T_{r_i}(Q_{j1}^*)).$$

Lastly, the card calculates the session key

$$SK = h(T_{r_i}(Q_{j1}^*)) || RSUID_j)$$

Step:5 $\quad MZ_s \rightarrow RSU_j : \{R_i\}$

Step: 6 the service $MZ_s$ checks whether the equality holds

$$R_i = h\left(RSUID_j \,||\, Q_{i1}\, \mathrm{P} Q_{j1}\, \mathrm{P} h\left(CID_i \,||\, B_j \,||\, t\right) || Y_1 \,\mathrm{P} T_{r_j}(Q_{i1}) T_{r_j}(Q_{i1})\right)$$

### 2.2.5. *Password Change Phase*

Then the smart card computes

$$d_1^{new} = d_1 \oplus h(pw_i \,||\, ID_i) \oplus h(pw_i^{new} \,||\, ID_i), d_2^{new} = h\left(ID_i \,||\, pw_i^{new}\right)$$

and replaces

$\{d_1, d_2\}$ with $\{d_1^{new}, d_2^{new}\}$.

## 3. Security Analysis

This protocol restrains the malicious attacks associated to the dynamic address configuration and hold the security.

### 3.1. *Address forgery Attack*

Using this protocol, a mobile vehicle attains a unique address with no replicate address detection. If any malicious mobile vehicle takeoffs the address of a real mobile vehicle to spread a wrong message (for instance, a replicate location detection message), firstly the message will be delivered to its nearby mobile vehicles. As the malicious mobile vehicle is not authenticated by the nearby vehicles, they reject the wrong message. Hence, the wrong message is restricted. Similarly, if a malicious mobile vehicle try to interact with a real mobile vehicle by taking an address, then the real mobile vehicle will obstruct the interaction with that vehicle since the malicious mobile vehicle is not authenticated [20].

### 3.2. *Address Exhaustion Attack*

When a malicious mobile vehicle request for an address from a nearby mobile vehicle or roadside, its request is declined as the malicious mobile vehicle or roadside is not authenticated.

#### 3.2.1. *Fake address conflict attack*

If an address conflict message is broadcasted by any malicious mobile vehicle, then this message is firstly received by the nearby mobile vehicles and they abandon this address conflict message as they cannot authenticate the malicious mobile vehicle. Furthermore, this protocol gains the address configuration with no replicate address detection, so to certify an assigned address' distinctiveness, it does not apply an address conflict message. So, this could be easily authenticated by any mobile vehicle that the received address conflict message is from a malicious mobile vehicle. Hence the streaming of a fake address message is avoided.

#### 3.2.2. *Replay Attack*

When a good vehicle X, sends an encrypted address request message, a malicious node interrupts it and again sends the same message to a vehicle Y to ask for an address to exhaust the address resources and occupy network resources. After Y receives the encrypted address request message, it can abandon this message as in this message the time stamp is expired. A malicious node interrupts the encrypted address response message from a good vehicle Y and again sends the same encrypted address response message to another node X and causes the address conflict. When X receives this encrypted address response message, it can discard this message because it cannot decrypt the encrypted address response message through its secret key. Hence, the replay attack is prevented.

## 4. Performance Evaluation

The previous methods [19, 21-26] are designated to make a comparison with our method because of the following reasons in Table 3, Table 4 and

**Table 3.**Compare reasons

| Methods | Storage overhead | Anonymity | Traceability | Authentication | Confidentiality | RSU-aided | Assumption |
|---|---|---|---|---|---|---|---|
| X.wang [19] | Low | ✓ | ✓ | ✓ | ✓ | ✗ | Normal |
| B.J Chang [21] | Low | ✓ | ✓ | ✓ | ✗ | ✗ | Normal |
| Chen, Y.-S [22] | High | ✓ | ✓ | ✓ | ✓ | ✗ | Normal |
| T.hwang,P. Gope[23] | Low | ✓ | ✗ | ✓ | ✗ | ✗ | Normal |
| R. Kumar, M. Dave [24] | Low | ✓ | ✓ | ✓ | ✗ | ✗ | Strong |
| Xiaonan [25] | High | ✓ | ✓ | ✓ | ✗ | ✗ | Normal |
| B. Ying [26] | Low | ✓ | ✓ | ✓ | ✗ | ✓ | Normal |
| Our proposed protocol | Low | ✓ | ✓ | ✓ | ✓ | ✓ | Normal |

**Table 4**.Efficiency comparisons of our scheme and the other related schemes [19, 22]

| | Our Scheme | X. Wang [19] | Chen, Y.-S[22] |
|---|---|---|---|
| Authorization phase | $4T_{xor}+3T_{hash}+2T_{asym}+2$ random number | $4T_{xor}+4T_{sym}+13T_{exp}+6$ random number | $2T_{asym}+1T_{hash}+2$ random numbers |
| Access service phase | $5T_{xor}+6T_{hash}+3T_{asym}+3$ random number | $4T_{exp}+4T_{sym}+4$ random numbers | $4T_{asym}+4T_{hash}+2T_{sym}+6$ random numbers |
| Computational cost | $\approx 500\ T_{sym}$ | $\approx 1028 T_{sym}$ | $\approx 602 T_{sym}$ |
| Computational time | 4.35 | 6.42 | 5.7 |

We have given a brief comparison of our scheme and the other prevailing schemes in terms of performance metrics which are important in road network authentication protocol, i.e., Authorization phase, Access service phase, computational cost, and computational time as shown (Table 4). In our experiments, to simulate a practical situation, the mobile vehicle are installed on two PCs in a local area network. In the registration phase, the presented scheme needs one hash operation to calculate $d_0=h(pw_i||ID_i) \oplus N$ on user side, needs one road side unit point multiplication operation and one modular inversion operation to obtain $d_1=c \oplus N$, $d_2=h(ID_i||pw_i)$, and needs one symmetric encryption operation to calculates $A_i=h(ID_i||r)$, $c=A_i \oplus d_0$ on the road side unit. While in authentication phase user side requires four scalar point multiplication operations to compute $Vrh(PW)P$ and $W=rR$; three hash operations to calculate $h(ST)$ , $u$ $Auth$ and $SK$; one symmetric encryption operation to find $Z$ and one symmetric decryption operation to decrypt message *Auths*. The MIX-ZONE SERVER requires two symmetric decryption operations to decrypt message $\{R_i\}$ and [1], one symmetric encryption operation to calculate *Auths*; one elliptic curve scalar point multiplication operations to calculate $V$; and two hash operations to get *Authu* and *SK*. Thus, the whole execution time of the computational time (s) is calculated as 4.35. As an evident of Table 4, our protocol has outperformed other protocols. It is renowned evidence that the symmetric encryption/decryption operation and the hash function operation hold the same computational cost. Consequently, the Computational cost of our proposed protocol, regarding the hash function operations, is more efficient as compared to other protocols.

## 5. Conclusion and Future work

In this paper, we put forward an Improved User's Authenticated Key Establishment Protocol for Road Networks. We establish a privacy preserving authentication protocol to verify the vehicle activities in a privacy preserving manner. We proposed a new method to adjust the vehicles speed which reduces the vehicle delay which might suffers from the network gap problem. We proposed a novel vehicle orientated privacy preserving technique for road network services that is also proficient of protecting the Road side unit from attacks. The road side unit can immediately decide which messages are trustworthy, and then send its opinion to neighbouring vehicles immediately. With the aid of road side unit server, vehicles can successfully increase decision accuracy while making decisions about the event messages.

The evaluation experiments based on NS-3 [18] to improve user's authenticated key establishment protocol have proved that our protocol has comparatively shorter response time, diminish cost, less packet lost information and better privacy preservation as related to those proposed in [19] and [20].

In future work, we will propose authentication with handover process using IPV6 protocol for 4G LTE cellular networks to reduce the computational cost, faster process, less steps (it is implementable with IPv6 protocol probably the quickest way is a listing of command, it will shows crypto) scheme by the future proposed scheme, the overall system complexity is able to be reduced.

## Acknowledgments

The authors would like to appreciate all anonymous reviewers for their insightful comments and constructive suggestions to polish this paper in high quality. This research was supported by national natural science foundation of China (61202376), China press and publication administration key laboratory project and shanghai key lab of modern optical system.

## Conflict of interest

The author declares that there is no conflict of interest regarding the publication of this paper.

.